\documentclass[twocolumn,showpacs,prl,superscriptaddress,floatfix]{revtex4}
\usepackage{amsmath,amssymb}
\usepackage{graphicx}

\def\(({\left(}
\def\)){\right)}                       
\def\[[{\left[}
\def\]]{\right]}

\newcommand{\be}{\begin{equation}}
\newcommand{\ee}{\end{equation}}
\newcommand{\bea}{\begin{eqnarray}}
\newcommand{\eea}{\end{eqnarray}}

\begin{document}
\title{Exhaustive enumeration unveils clustering and freezing in random 3-SAT}

\author{John Ardelius}
\affiliation{Swedish Institute of Computer Science SICS, SE-164 24, Kista, Sweden}
\email{john@sics.se}
\author{Lenka Zdeborov\'a}
\affiliation{ Universit\'e Paris-Sud, LPTMS, UMR8626,  B\^{a}t.~100, Universit\'e
Paris-Sud 91405 Orsay cedex}
\affiliation{CNRS, LPTMS, UMR8626, B\^{a}t.~100, Universit\'e Paris-Sud 91405 Orsay cedex}
\email{zdeborov@lptms.u-psud.fr}

\begin{abstract}

We study geometrical properties of the complete set of solutions of the random 3-satisfiability problem. We show that even for moderate system sizes the number of clusters corresponds surprisingly well with the theoretic asymptotic prediction. We locate the freezing transition in the space of solutions which has been conjectured to be relevant in explaining the onset of computational hardness in random constraint satisfaction problems. 

\end{abstract}

\date{\today}
\maketitle

Satisfiability (SAT) is one of the most important problems in theoretical computer science. It was the first problem shown to be NP-complete \cite{Cook71,Papadimitriou94}, and it is of central relevance in various  practical applications, including artificial intelligence, planning, hardware and electronic design, automation, verification and more. It can thus be pictorially thought of as the Ising model of computer science. 
Ensembles of randomly generated SAT instances emerged in computer science as a way of evaluating algorithmic performance and addressing questions regarding the average case complexity.

An instance of random $K$-SAT problem consists of $N$ Boolean variables and $M$ clauses.
Each clause contains a subset of $K$ distinct variables chosen uniformly at random, and each clause forbids one random assignment of the $K$ variables out of the $2^K$ possible ones. 
The problem is \textit{satisfiable} if there exists a variable assignment that simultaneously satisfies all clauses and we call such an assignments a {\it solution} to the problem. When the density of constraints $\alpha=M/N$ is increased, the formulas become less likely to be satisfiable. In the thermodynamical limit there is a sharp transition from a phase in which the formulas are almost surely satisfiable to a phase where they are almost surely unsatisfiable. The existence of this transition is partly established rigorously \cite{Friedgut99}. It is also a well known empirical result that the hardest instances are found near to this threshold \cite{CheesemanKanefsky91,MitchellSelman92,KirkpatrickSelman94}.

Random $K$-SAT has attracted interest of statistical physicists because of its equivalence to mean field spin glasses \cite{MezardParisi87b}. Indeed, the problem can be rephrased as minimizing a spin glass-like energy function which counts the number of violated clauses. 
The results and insights coming from this equivalence are remarkable. The satisfiability threshold and other phase transitions in the structure of solutions are described in \cite{BiroliMonasson00,MezardParisi02,KrzakalaMontanari06}.  In particular, it was shown that for $K\ge 3$ the space of solutions for highly constrained but still satisfiable instances splits into exponentially many clusters and in some cases this clustering has been rigorously confirmed \cite{MezardMora05,AchlioptasRicci06}. The so-called freezing of variables in clusters is another rich concept studied recently \cite{Semerjian07,ZdeborovaKrzakala07}.
However, a detailed understanding of how the clustering or freezing of solutions affects the average computational hardness is still one of the most interesting open problems in the field.

Since the exact statistical physics solution of the random satisfiability problem appeared \cite{MezardParisi02,MezardZecchina02} dozens of directly related articles followed. Mathematicians and computer scientist nowadays regard these analytical works as a rich source of results which are mostly unaccessible to the current probabilistic methods. Yet none of these works tried to compare the analytical asymptotic predictions to numerical simulations on a quantitative level and numerical investigations mostly concentrated on  performance analysis of satisfiability solvers. Therefore the relevance of the asymptotic predictions for systems of practical sizes, which in computer science are not at the scale of the Avogadro number, remained almost untouched. 
Our letter aims at filling this gap and to encouraging further investigation in this direction. We use conceptually relatively simple numerical techniques and yet obtain nontrivial results.  We present two of our most interesting findings. The first is a quantitative comparison between the number of clusters of solutions (glassy states) and its analytical prediction \cite{MezardParisi02,MezardZecchina02,MontanariParisi04,MertensMezard06}. The second is the location of the freezing transition which was recently suggested to be responsible for computational hardness of the random satisfiability problem \cite{ZdeborovaKrzakala07,KrzakalaKurchan07,DallAstaRamezanpour08}, but not yet computed in the 3-SAT problem.

\paragraph*{Clustering and freezing ---}
In physics of glassy systems, clusters correspond to pure thermodynamical states and are being described in the literature about glasses and spin glasses for more than one quarter of a century \cite{MezardParisi87b}. A formal definition of clusters in $K$-SAT as extremal Gibbs measures was given recently in \cite{KrzakalaMontanari06}. We will refer to these as the {\it cavity-clusters}. It is not known, however, how to adapt this definition to instances of finite size.  In this work, we define clusters as connected components in a graph where each solution is a vertex and where edges connect solutions that differ in only one variable \footnote{Another common choice of distance between neighborhooding solutions is "any sub-extensive distance". This choice is, however, problematic for finite-size instances and lacks the important property that all the solutions in a cluster have the same whitening core. In some other models this definition is to be modified, e.g., in the 1-in-$K$ SAT the minimal distance between solutions is 2.}.
This definition is applicable to any finite instance of the $K$-SAT problem. It is most likely not strictly equivalent to the definition of the cavity-clusters, yet it reproduces many of their properties.

In order to shed light on the relation between cavity-clusters and connected-component clusters we now introduce the procedure called {\it whitening} and the concept of frozen variables. \textit{Whitening of a solution} in $K$-SAT is defined in the following way \cite{Parisi02b}: start with the solution, assign iteratively a "$\ast$" (joker) to variables which belong only to clauses which are already satisfied by another variable or already contain a $\ast$  variable~\footnote{In a general constraint satisfaction problem the whitening mush be defined via the warning propagation.}. Whitening is in the literature referred also as peeling \cite{ManevaMossel05} or coarsening \cite{AchlioptasRicci06}.
The fixed point of this procedure is called a {\it whitening-core}, it is also referred as core \cite{ManevaMossel05,AchlioptasRicci06}, or true cover \cite{KrocSabharwal07}.
A variable is said to be \textit{frozen} in a set of solutions if it takes only one value (either 0 or 1) in all the solutions in the set. Note that if the satisfiability threshold is sharp there cannot be a finite fraction of variables frozen in all the solutions in the satisfiable region \cite{MonassonZecchina99}. On the other hand, variables might be frozen in the individual clusters. According to the cavity method \cite{MezardParisi01,MezardParisi03} this is indeed the case and freezing of clusters have been studied in \cite{ZdeborovaKrzakala07,Semerjian07,MontanariRicci08}.

According to the cavity method \cite{MezardParisi01,MezardParisi03} there is a deep connection between frozen variables and the whitening-core: if the one-step replica symmetric solution is correct then on large typical instances the set of frozen variables in the cavity-cluster and the non-$\ast$ part of the whitening core are identical \cite{MezardParisi02}. Thus the whitening cores of all solutions belonging to one cavity-cluster are identical.  
This also holds for the connected-component clusters: Indeed, two solutions that differ in a single variable have the same whitening core  
since the whitening can be started from that specific variable~\footnote{ 
We have not found any general argument why two different connected-component clusters could not have the same non-all-$\ast$ whitening core, but we have not observed any such case in our data.}. 
Further, variables belonging to the whitening core must be frozen in the connected-component cluster, the opposite implication is in general not true~\footnote{We are aware of one property which the cavity-clusters might have but which cannot be reproduced asymptotically with the connected-component clusters, namely a purely entropic separation.}. 

Two additional remarks about clusters are important. First, whitening cores are sometimes wrongly identified with clusters. In part of the clustered phase almost all solutions belong to soft (unfrozen) cavity-clusters \cite{ZdeborovaKrzakala07,MontanariRicci08}. In particular in 3-SAT this seems to be the case at least up to constraint density $\alpha=4.25$ \cite{Fede}. Second, it seems that all known heuristic algorithms need an exponential time to find solutions with a non-trivial (not all-$\ast$) whitening cores, see e.g. \cite{ManevaMossel05,SeitzAlava05,ZdeborovaKrzakala07}. This motivates our  study of the \textit{freezing transition}, $\alpha_f$. It is defined as the smallest density of constraints $\alpha$ such that all solutions belong to frozen clusters, i.e., their whitening core is not made from all-$\ast$. We use the whitening core instead of the real set of frozen variables, because in small instances there are almost always at least few frozen variables. 
Existence of the frozen phase was proven in the thermodynamical limit for $K\ge 9$ near to the satisfiability threshold in \cite{AchlioptasRicci06}. Several theoretical investigations of a related rigidity transition, where clusters which contain almost all the solution become frozen, can be found in \cite{Semerjian07,ZdeborovaKrzakala07,MontanariRicci08}. But as long as soft clusters exist some algorithms may be able to find them, as shown in \cite{DallAstaRamezanpour08}. A~related numerical study \cite{KrocSabharwal07} investigates the size dependence of the fraction of frozen solutions at $\alpha=4.20<\alpha_f$.  

\paragraph*{The complexity function ---}

We generate instances of random 3-SAT problems with $N$ variables and $M$ clauses using the {\tt makewff} program \cite{makewff}. 
The number of solutions is then calculated using the exhaustive search method {\tt relsat} \cite{BayardoPehousek00} and the complete set of solutions is clustered through {\it breath first search}. This works as follows: We order
the ${\cal N}$ solutions in binary lexical order. Further, for all the solutions we generate all the $N$ neighboring configurations, search them in the list and if found concatenate the two in the same cluster resulting in an algorithmic complexity of $O({\cal N} \log^2{\cal N})$, considering that $\log{\cal N} \approx N$.

\begin{figure}[!ht]
  \resizebox{8.9cm}{!}{\includegraphics[angle=270]{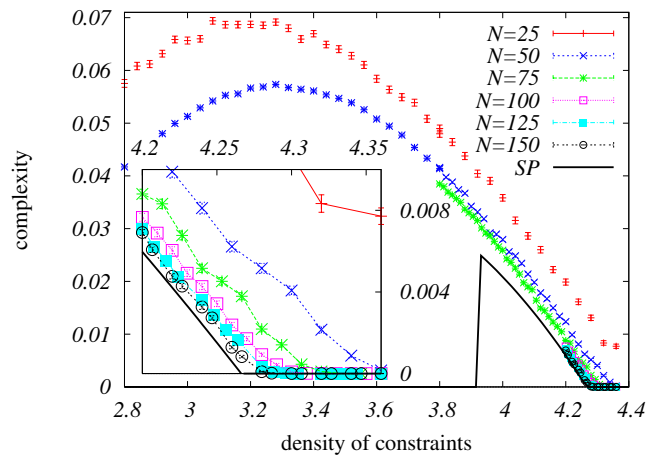}}
  \caption{\label{cplx} The average complexity function, logarithm of the number of connected-component clusters divided by $N$, for different system sizes compared to the asymptotic prediction \cite{MezardZecchina02,MertensMezard06}. Note that the numerical curves will continue to much lower values of $\alpha$ than plotted.}
\end{figure}

In order to obtain information about clusters in a typical formula with $N$ variables and $M$ clauses, we count the number of solutions in $A=999$ such random formulas and select the median instance in terms of number of solutions on which we then count the number of clusters~${\cal S}$. This is repeated $B=1000$ times. The complexity function $\Sigma(N) = \langle \log{\cal S} \rangle/N$ is then computed as average of the logarithm of the number of clusters divided by system size~$N$. If the median instance is unsatisfiable it contributes a zero value to the average, this does not have influence of the asymptotic value. Taking the median has two important advantages, first we avoid  rare formulas with very many solutions which are numerically intractable, second the complexity converges very fast to zero in the unsatisfiable region. 
The result is plotted in Fig.~\ref{cplx} and compared with the asymptotic complexity function computed from the survey propagation equations, which in 3-SAT gives a non-zero result for $\alpha>3.92$ \cite{MezardZecchina02,MertensMezard06}. The agreement is remarkably good, in particular around the satisfiability threshold $\alpha_s=4.267$~\cite{MezardParisi02,MertensMezard06}. 

It was discussed in \cite{KrzakalaMontanari06}, and shown numerically in \cite{KrocSabharwal07}, that clusters exist even for $\alpha<3.92$.
We indeed do not see anything particular happening at $\alpha=3.92$.  Below the clustering transition, $\alpha<3.86$, however, the largest cavity-cluster should contain almost all the solutions \cite{KrzakalaMontanari06}. We see a corresponding trend in the average fraction of solutions covered by the largest cluster in our data. It should also be mentioned that the survey propagation prediction is believed to be exact only for $\alpha>4.15$ \cite{MontanariParisi04}.

\begin{figure}[!ht]
 \resizebox{8.7cm}{!}{\includegraphics[angle=270]{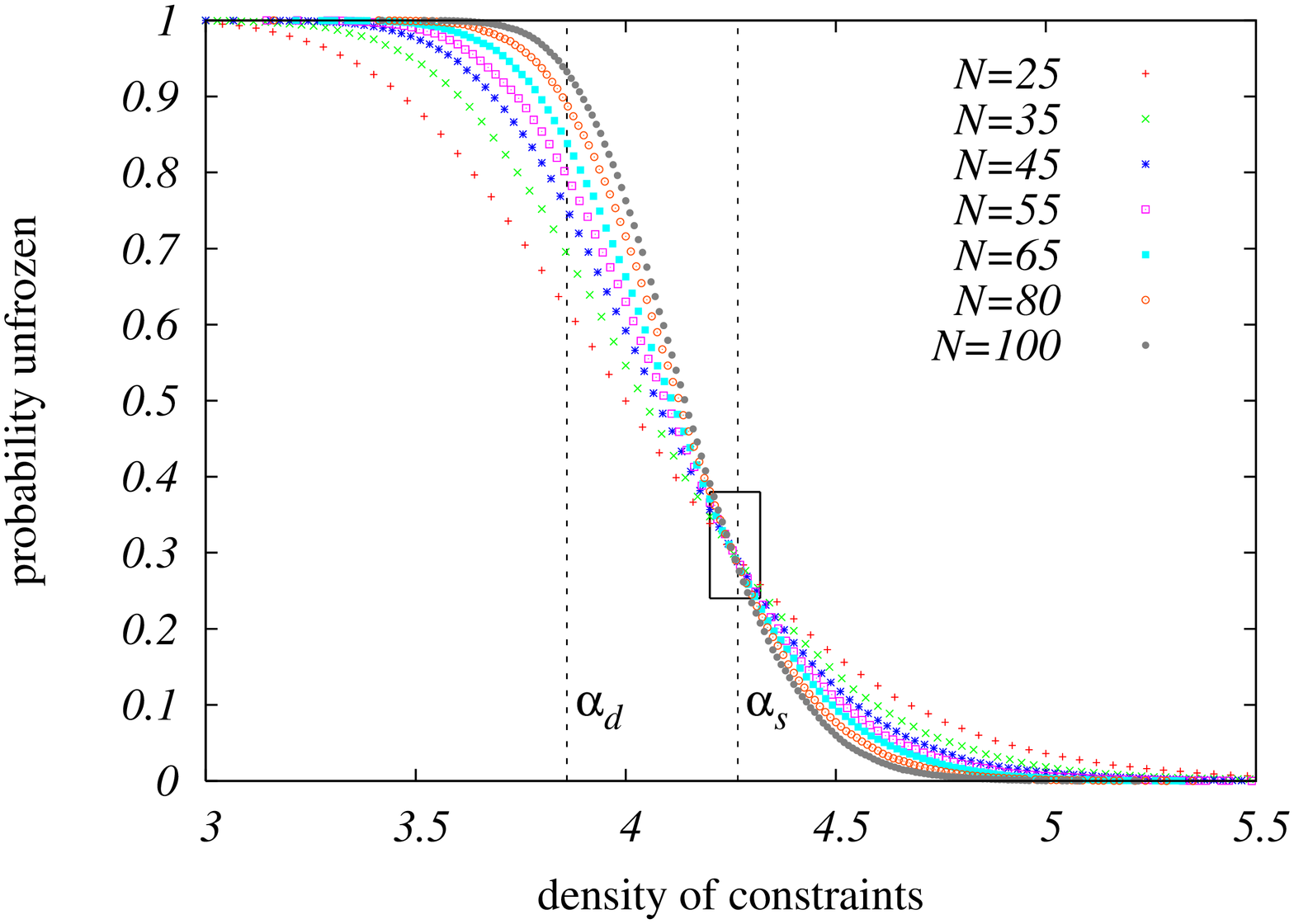}}
 \resizebox{8.7cm}{!}{\includegraphics[angle=270]{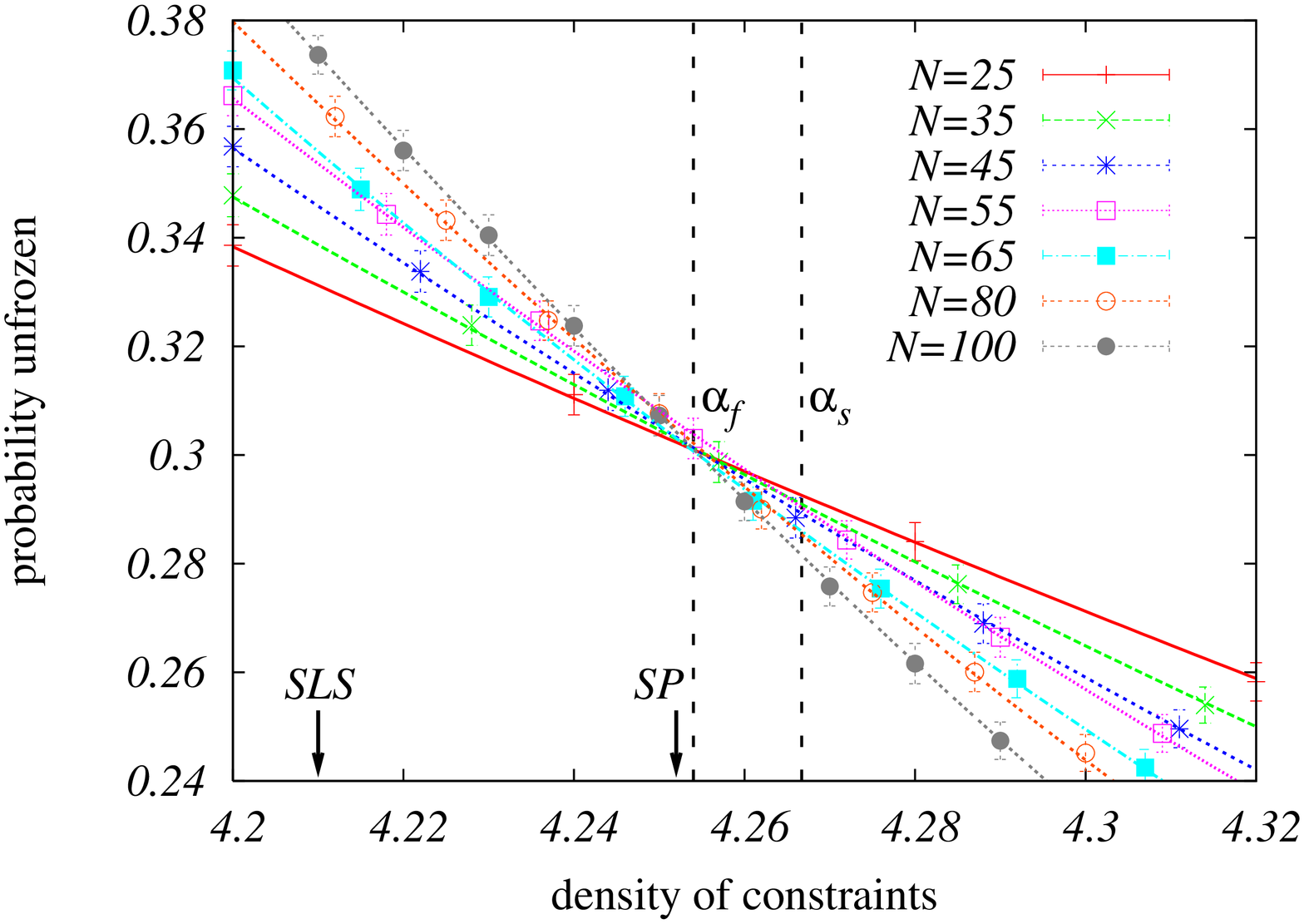}}
\caption{\label{fig:rig} Top: Probability that there exists an unfrozen solution as a function of the constraint density $\alpha$ for different system sizes. The clustering \cite{KrzakalaMontanari06} and satisfiability \cite{MezardParisi02} transitions are marked for comparison. Bottom: A 1:20 zoom on the critical (crossing) point, our estimate for the freezing transition is $\alpha_f=4.254\pm 0.009$. The curves are cubic fits in the interval $(4,4.4)$. The arrows represent estimates of the limits of performance of the best known stochastic local search \cite{SeitzAlava05,ArdeliusAurell06} and survey propagation \cite{Parisi03,ChavasFurtlehner05} algorithms.
\vspace{-0.2cm}}
\end{figure}

\paragraph{The freezing transition ---} In order to determine the freezing transition we start with a formula of $N$ variables and all possible clauses, and remove the clauses one by one independently at random. We mark the number of clauses $M_s$ where the formula becomes satisfiable as well as the number of clauses $M_f \le M_s$ where at least one solution starts to have an all-$\ast$ whitening core. We repeat $B$-times ($B=2\cdot10^4$ in Fig.~\ref{fig:rig}) and compute the probabilities that a formula of $M$ clauses is satisfiable $P_s(\alpha,N)$, and unfrozen $P_f(\alpha,N)$, respectively. 
Due to memory limitation we can treat only instances which have less than $5\cdot10^7$ solutions which limits us to system sizes $N\le 100$. Our results for the satisfiability threshold are consistent
with previous studies in \cite{KirkpatrickSelman94,MonassonZecchina99,MonassonZecchina99b}. The probability of being unfrozen, $P_f(\alpha,N)$, is shown in Fig.~\ref{fig:rig}. 

It is tempting to perform a scaling analysis as has been done in \cite{KirkpatrickSelman94,MonassonZecchina99,MonassonZecchina99b} for the satisfiability threshold. The critical exponent related to the width of the scaling window was defined via rescaling of variable $\alpha$ as $N^{1/\nu_s}(1-\alpha/\alpha_s(K,N))$. Note, however, that the estimate $\nu_s=1.5\pm 0.1$ for 3-SAT provided in \cite{MonassonZecchina99b} is not the correct asymptotic value. It was proven in \cite{Wilson02} that $\nu_s\ge 2$. Indeed it was shown numerically in \cite{LeoneRicci01} that a crossover exists at sizes of order $N \approx 10^4$ in the related XOR-SAT problem. A similar situation happens for the scaling of the freezing transition, $P_f(\alpha,N)$, as the proof of \cite{Wilson02} applies also here~\footnote{Theorem 1 of \cite{Wilson02} applies to the freezing property where the bystander are clauses containing two leaves.}. It would be interesting to investigate the scaling behavior on an ensemble of instances where results of \cite{Wilson02} do not apply. Here we concentrate instead on the estimation of the critical point, which we presume not to be influenced by the crossover in the scaling. We are in a much more convenient situation than for the satisfiability transition. The crossing point for the functions $P_f(\alpha,N)$ of different system sizes seems not to depend on $N$, while for the satisfiability transition its size dependence is very strong \cite{MonassonZecchina99b}.

We determine the value of the freezing transition as $\alpha_f=4.254\pm 0.009$, which is extremely close to the satisfiability threshold $\alpha_s=4.267$ \cite{MezardZecchina02}.
Analytical study suggests $\alpha_f>4.25$ \cite{Fede}. We expect the two transitions to be separated $\alpha_f<\alpha_s$ \cite{AchlioptasRicci06,ZdeborovaKrzakala07,MontanariRicci08}, and Fig. \ref{fig:rig} suggests so but it is on the border of statistical significance. However, the main motivation to study the freezing transition is its potential connection to the onset of algorithmical hardness \cite{ZdeborovaKrzakala07,KrzakalaKurchan07,DallAstaRamezanpour08}. We thus compare its value with the estimates of performance of the best algorithms known for random 3-SAT. The leading stochastic local search algorithms work in linear time up to $\alpha=4.21$ \cite{SeitzAlava05,ArdeliusAurell06}. The survey propagation (SP) decimation was estimated to work up to $\alpha=4.252$ \cite{Parisi03}, the same point
was determined as the limit of the SP reinforcement \cite{ChavasFurtlehner05}. The agreement between our location of the freezing transition and the performance of SP supports strongly the conjecture that the frozen phase is hard for any known algorithm. In random 3-SAT this region is very narrow, in contrast to the situation in $K\ge 9$ SAT~\cite{AchlioptasRicci06}. 

\paragraph{Discussion ---}

The main contribution of this work is the demonstration that the asymptotic predictions coming from the statistical physics analysis are relevant even for instances of very moderate size. In particular, 
we presented a numerical comparison between the number of connected-component clusters and the asymptotic prediction for the complexity function in random 3-SAT and obtain a remarkably good agreement. Furthermore, we estimate the location of the freezing transition at $\alpha_f=4.254$, which is consistent with the performance threshold of the best known algorithms. 
We also show that exhaustive enumeration, despite its current size limitations, is a powerful tool to study random optimization problems: indeed the knowledge
of the complete set of solutions allows to tackle questions that are complementary to those answered by classical Monte-Carlo methods.

The definitions of clusters and the whitening core, that we adopted, is applicable to any instance of the satisfiability problem. As such, they offer an interesting direction for future research of real-world $K$-SAT instances.
In addition, we observe that the properties related to clustering are less sensitive to finite-size effects than the ones related to the solutions themselves. This is interesting and certainly worth further investigations. Future work could also cover 2-SAT, where the solutions are much more numerous even for very small system sizes, or $K$-SAT with $K>3$, where larger formulas will be needed to investigate the relevant regions, however, the freezing transition is more separated from the satisfiability when $K$ grows. The numerical location of the clustering and condensation transitions \cite{KrzakalaMontanari06} is also of interest.

\paragraph{Acknowledgment ---}

We thank Stephan Mertens for sharing the data from \cite{MertensMezard06}. We also gratefully acknowledge Thomas Joerg, Florent Krzakala, Marc M\'ezard and Federico Ricci-Tersenghi for many precious discussions and comments. This work was partially supported by FP6 program EVERGROW. J.A. thanks KITPC-CAS for hospitality.

\bibliography{myentries}

\begin{thebibliography}{36}
\expandafter\ifx\csname natexlab\endcsname\relax\def\natexlab#1{#1}\fi
\expandafter\ifx\csname bibnamefont\endcsname\relax
  \def\bibnamefont#1{#1}\fi
\expandafter\ifx\csname bibfnamefont\endcsname\relax
  \def\bibfnamefont#1{#1}\fi
\expandafter\ifx\csname citenamefont\endcsname\relax
  \def\citenamefont#1{#1}\fi
\expandafter\ifx\csname url\endcsname\relax
  \def\url#1{\texttt{#1}}\fi
\expandafter\ifx\csname urlprefix\endcsname\relax\def\urlprefix{URL }\fi
\providecommand{\bibinfo}[2]{#2}
\providecommand{\eprint}[2][]{\url{#2}}

\bibitem[{\citenamefont{Cook}(1971)}]{Cook71}
\bibinfo{author}{\bibfnamefont{S.~A.} \bibnamefont{Cook}}, in
  \emph{\bibinfo{booktitle}{Proc. 3rd STOC}} (\bibinfo{publisher}{ACM},
  \bibinfo{address}{New York, NY, USA}, \bibinfo{year}{1971}), pp.
  \bibinfo{pages}{151--158}.

\bibitem[{\citenamefont{Papadimitriou}(1994)}]{Papadimitriou94}
\bibinfo{author}{\bibfnamefont{C.~H.} \bibnamefont{Papadimitriou}},
  \emph{\bibinfo{title}{Computational complexity}}
  (\bibinfo{publisher}{Addison-Wesley}, \bibinfo{year}{1994}).

\bibitem[{\citenamefont{Friedgut}(1999)}]{Friedgut99}
\bibinfo{author}{\bibfnamefont{E.}~\bibnamefont{Friedgut}},
  \bibinfo{journal}{J. Amer. Math. Soc.} \textbf{\bibinfo{volume}{12}}
  (\bibinfo{year}{1999}).

\bibitem[{\citenamefont{Cheeseman et~al.}(1991)\citenamefont{Cheeseman,
  Kanefsky, and Taylor}}]{CheesemanKanefsky91}
\bibinfo{author}{\bibfnamefont{P.}~\bibnamefont{Cheeseman}},
  \bibinfo{author}{\bibfnamefont{B.}~\bibnamefont{Kanefsky}}, \bibnamefont{and}
  \bibinfo{author}{\bibfnamefont{W.~M.} \bibnamefont{Taylor}}, in
  \emph{\bibinfo{booktitle}{Proc. 12th IJCAI}} (\bibinfo{publisher}{Morgan
  Kaufmann}, \bibinfo{address}{San Mateo, CA, USA}, \bibinfo{year}{1991}), pp.
  \bibinfo{pages}{331--337}.

\bibitem[{\citenamefont{Mitchell et~al.}(1992)\citenamefont{Mitchell, Selman,
  and Levesque}}]{MitchellSelman92}
\bibinfo{author}{\bibfnamefont{D.~G.} \bibnamefont{Mitchell}},
  \bibinfo{author}{\bibfnamefont{B.}~\bibnamefont{Selman}}, \bibnamefont{and}
  \bibinfo{author}{\bibfnamefont{H.~J.} \bibnamefont{Levesque}}, in
  \emph{\bibinfo{booktitle}{Proc. 10th AAAI}} (\bibinfo{publisher}{AAAI Press},
  \bibinfo{address}{Menlo Park, California}, \bibinfo{year}{1992}), pp.
  \bibinfo{pages}{459--465}.

\bibitem[{\citenamefont{Kirkpatrick and Selman}(1994)}]{KirkpatrickSelman94}
\bibinfo{author}{\bibfnamefont{S.}~\bibnamefont{Kirkpatrick}} \bibnamefont{and}
  \bibinfo{author}{\bibfnamefont{B.}~\bibnamefont{Selman}},
  \bibinfo{journal}{Science} \textbf{\bibinfo{volume}{264}},
  \bibinfo{pages}{1297} (\bibinfo{year}{1994}).

\bibitem[{\citenamefont{M{\'e}zard et~al.}(1987)\citenamefont{M{\'e}zard,
  Parisi, and Virasoro}}]{MezardParisi87b}
\bibinfo{author}{\bibfnamefont{M.}~\bibnamefont{M{\'e}zard}},
  \bibinfo{author}{\bibfnamefont{G.}~\bibnamefont{Parisi}}, \bibnamefont{and}
  \bibinfo{author}{\bibfnamefont{M.~A.} \bibnamefont{Virasoro}},
  \emph{\bibinfo{title}{Spin-Glass Theory and Beyond}},
  vol.~\bibinfo{volume}{9} of \emph{\bibinfo{series}{Lecture Notes in Physics}}
  (\bibinfo{publisher}{World Scientific}, \bibinfo{address}{Singapore},
  \bibinfo{year}{1987}).

\bibitem[{\citenamefont{Biroli et~al.}(2000)\citenamefont{Biroli, Monasson, and
  Weigt}}]{BiroliMonasson00}
\bibinfo{author}{\bibfnamefont{G.}~\bibnamefont{Biroli}},
  \bibinfo{author}{\bibfnamefont{R.}~\bibnamefont{Monasson}}, \bibnamefont{and}
  \bibinfo{author}{\bibfnamefont{M.}~\bibnamefont{Weigt}},
  \bibinfo{journal}{Eur. Phys. J. B} \textbf{\bibinfo{volume}{14}},
  \bibinfo{pages}{551} (\bibinfo{year}{2000}).

\bibitem[{\citenamefont{M{\'e}zard et~al.}(2002)\citenamefont{M{\'e}zard,
  Parisi, and Zecchina}}]{MezardParisi02}
\bibinfo{author}{\bibfnamefont{M.}~\bibnamefont{M{\'e}zard}},
  \bibinfo{author}{\bibfnamefont{G.}~\bibnamefont{Parisi}}, \bibnamefont{and}
  \bibinfo{author}{\bibfnamefont{R.}~\bibnamefont{Zecchina}},
  \bibinfo{journal}{Science} \textbf{\bibinfo{volume}{297}},
  \bibinfo{pages}{812} (\bibinfo{year}{2002}).

\bibitem[{\citenamefont{Krzakala et~al.}(2007)\citenamefont{Krzakala,
  Montanari, Ricci-Tersenghi, Semerjian, and
  Zdeborov{\'a}}}]{KrzakalaMontanari06}
\bibinfo{author}{\bibfnamefont{F.}~\bibnamefont{Krzakala}},
  \bibinfo{author}{\bibfnamefont{A.}~\bibnamefont{Montanari}},
  \bibinfo{author}{\bibfnamefont{F.}~\bibnamefont{Ricci-Tersenghi}},
  \bibinfo{author}{\bibfnamefont{G.}~\bibnamefont{Semerjian}},
  \bibnamefont{and}
  \bibinfo{author}{\bibfnamefont{L.}~\bibnamefont{Zdeborov{\'a}}},
  \bibinfo{journal}{Proc. Natl. Acad. Sci.} \textbf{\bibinfo{volume}{104}},
  \bibinfo{pages}{10318} (\bibinfo{year}{2007}).

\bibitem[{\citenamefont{M\'ezard et~al.}(2005)\citenamefont{M\'ezard, Mora, and
  Zecchina}}]{MezardMora05}
\bibinfo{author}{\bibfnamefont{M.}~\bibnamefont{M\'ezard}},
  \bibinfo{author}{\bibfnamefont{T.}~\bibnamefont{Mora}}, \bibnamefont{and}
  \bibinfo{author}{\bibfnamefont{R.}~\bibnamefont{Zecchina}},
  \bibinfo{journal}{Phys. Rev. Lett.} \textbf{\bibinfo{volume}{94}},
  \bibinfo{pages}{197205} (\bibinfo{year}{2005}).

\bibitem[{\citenamefont{Achlioptas and
  Ricci-Tersenghi}(2006)}]{AchlioptasRicci06}
\bibinfo{author}{\bibfnamefont{D.}~\bibnamefont{Achlioptas}} \bibnamefont{and}
  \bibinfo{author}{\bibfnamefont{F.}~\bibnamefont{Ricci-Tersenghi}}, in
  \emph{\bibinfo{booktitle}{Proc. of 38th STOC}} (\bibinfo{publisher}{ACM},
  \bibinfo{address}{New York, NY, USA}, \bibinfo{year}{2006}), pp.
  \bibinfo{pages}{130--139}.

\bibitem[{\citenamefont{Semerjian}(2008)}]{Semerjian07}
\bibinfo{author}{\bibfnamefont{G.}~\bibnamefont{Semerjian}},
  \bibinfo{journal}{J. Stat. Phys.} \textbf{\bibinfo{volume}{130}},
  \bibinfo{pages}{251} (\bibinfo{year}{2008}).

\bibitem[{\citenamefont{Zdeborov{\'a} and
  Krzakala}(2007)}]{ZdeborovaKrzakala07}
\bibinfo{author}{\bibfnamefont{L.}~\bibnamefont{Zdeborov{\'a}}}
  \bibnamefont{and} \bibinfo{author}{\bibfnamefont{F.}~\bibnamefont{Krzakala}},
  \bibinfo{journal}{Phys. Rev. E} \textbf{\bibinfo{volume}{76}},
  \bibinfo{pages}{031131} (\bibinfo{year}{2007}).

\bibitem[{\citenamefont{M{\'e}zard and Zecchina}(2002)}]{MezardZecchina02}
\bibinfo{author}{\bibfnamefont{M.}~\bibnamefont{M{\'e}zard}} \bibnamefont{and}
  \bibinfo{author}{\bibfnamefont{R.}~\bibnamefont{Zecchina}},
  \bibinfo{journal}{Phys. Rev. E} \textbf{\bibinfo{volume}{66}},
  \bibinfo{pages}{056126} (\bibinfo{year}{2002}).

\bibitem[{\citenamefont{Montanari et~al.}(2004)\citenamefont{Montanari, Parisi,
  and Ricci-Tersenghi}}]{MontanariParisi04}
\bibinfo{author}{\bibfnamefont{A.}~\bibnamefont{Montanari}},
  \bibinfo{author}{\bibfnamefont{G.}~\bibnamefont{Parisi}}, \bibnamefont{and}
  \bibinfo{author}{\bibfnamefont{F.}~\bibnamefont{Ricci-Tersenghi}},
  \bibinfo{journal}{J. Phys. A} \textbf{\bibinfo{volume}{37}},
  \bibinfo{pages}{2073} (\bibinfo{year}{2004}).

\bibitem[{\citenamefont{Mertens et~al.}(2006)\citenamefont{Mertens, M\'ezard,
  and Zecchina}}]{MertensMezard06}
\bibinfo{author}{\bibfnamefont{S.}~\bibnamefont{Mertens}},
  \bibinfo{author}{\bibfnamefont{M.}~\bibnamefont{M\'ezard}}, \bibnamefont{and}
  \bibinfo{author}{\bibfnamefont{R.}~\bibnamefont{Zecchina}},
  \bibinfo{journal}{Rand. Struct. Algo.} \textbf{\bibinfo{volume}{28}},
  \bibinfo{pages}{340} (\bibinfo{year}{2006}).

\bibitem[{\citenamefont{Krzakala and Kurchan}(2007)}]{KrzakalaKurchan07}
\bibinfo{author}{\bibfnamefont{F.}~\bibnamefont{Krzakala}} \bibnamefont{and}
  \bibinfo{author}{\bibfnamefont{J.}~\bibnamefont{Kurchan}},
  \bibinfo{journal}{Phys. Rev. E} \textbf{\bibinfo{volume}{76}},
  \bibinfo{pages}{021122} (\bibinfo{year}{2007}).

\bibitem[{\citenamefont{Dall'Asta et~al.}(2008)\citenamefont{Dall'Asta,
  Ramezanpour, and Zecchina}}]{DallAstaRamezanpour08}
\bibinfo{author}{\bibfnamefont{L.}~\bibnamefont{Dall'Asta}},
  \bibinfo{author}{\bibfnamefont{A.}~\bibnamefont{Ramezanpour}},
  \bibnamefont{and} \bibinfo{author}{\bibfnamefont{R.}~\bibnamefont{Zecchina}},
  \bibinfo{journal}{Phys. Rev. E} \textbf{\bibinfo{volume}{77}}
  (\bibinfo{year}{2008}).

\bibitem[{\citenamefont{Parisi}(2002)}]{Parisi02b}
\bibinfo{author}{\bibfnamefont{G.}~\bibnamefont{Parisi}}
  (\bibinfo{year}{2002}), \bibinfo{note}{arXiv:cs.CC/0212047}.

\bibitem[{\citenamefont{Maneva et~al.}(2007)\citenamefont{Maneva, Mossel, and
  Wainwright}}]{ManevaMossel05}
\bibinfo{author}{\bibfnamefont{E.~N.} \bibnamefont{Maneva}},
  \bibinfo{author}{\bibfnamefont{E.}~\bibnamefont{Mossel}}, \bibnamefont{and}
  \bibinfo{author}{\bibfnamefont{M.~J.} \bibnamefont{Wainwright}},
  \bibinfo{journal}{J. ACM} \textbf{\bibinfo{volume}{54}}
  (\bibinfo{year}{2007}).

\bibitem[{\citenamefont{Kroc et~al.}(2007)\citenamefont{Kroc, Sabharwal, and
  Selman}}]{KrocSabharwal07}
\bibinfo{author}{\bibfnamefont{L.}~\bibnamefont{Kroc}},
  \bibinfo{author}{\bibfnamefont{A.}~\bibnamefont{Sabharwal}},
  \bibnamefont{and} \bibinfo{author}{\bibfnamefont{B.}~\bibnamefont{Selman}},
  in \emph{\bibinfo{booktitle}{Proc. of 23rd AUAI}} (\bibinfo{publisher}{AUAI
  Press}, \bibinfo{address}{Arlington, Virginia, USA}, \bibinfo{year}{2007}),
  pp. \bibinfo{pages}{217--226}.

\bibitem[{\citenamefont{Monasson
  et~al.}(1999{\natexlab{a}})\citenamefont{Monasson, Zecchina, Kirkpatrick,
  Selman, and Troyansky}}]{MonassonZecchina99}
\bibinfo{author}{\bibfnamefont{R.}~\bibnamefont{Monasson}},
  \bibinfo{author}{\bibfnamefont{R.}~\bibnamefont{Zecchina}},
  \bibinfo{author}{\bibfnamefont{S.}~\bibnamefont{Kirkpatrick}},
  \bibinfo{author}{\bibfnamefont{B.}~\bibnamefont{Selman}}, \bibnamefont{and}
  \bibinfo{author}{\bibfnamefont{L.}~\bibnamefont{Troyansky}},
  \bibinfo{journal}{Nature} \textbf{\bibinfo{volume}{400}},
  \bibinfo{pages}{133} (\bibinfo{year}{1999}{\natexlab{a}}).

\bibitem[{\citenamefont{M{\'e}zard and Parisi}(2003)}]{MezardParisi03}
\bibinfo{author}{\bibfnamefont{M.}~\bibnamefont{M{\'e}zard}} \bibnamefont{and}
  \bibinfo{author}{\bibfnamefont{G.}~\bibnamefont{Parisi}},
  \bibinfo{journal}{J. Stat. Phys.} \textbf{\bibinfo{volume}{111}},
  \bibinfo{pages}{1} (\bibinfo{year}{2003}).

\bibitem[{\citenamefont{M{\'e}zard and Parisi}(2001)}]{MezardParisi01}
\bibinfo{author}{\bibfnamefont{M.}~\bibnamefont{M{\'e}zard}} \bibnamefont{and}
  \bibinfo{author}{\bibfnamefont{G.}~\bibnamefont{Parisi}},
  \bibinfo{journal}{Eur. Phys. J. B} \textbf{\bibinfo{volume}{20}},
  \bibinfo{pages}{217} (\bibinfo{year}{2001}).

\bibitem[{\citenamefont{Montanari et~al.}(2008)\citenamefont{Montanari,
  Ricci-Tersenghi, and Semerjian}}]{MontanariRicci08}
\bibinfo{author}{\bibfnamefont{A.}~\bibnamefont{Montanari}},
  \bibinfo{author}{\bibfnamefont{F.}~\bibnamefont{Ricci-Tersenghi}},
  \bibnamefont{and}
  \bibinfo{author}{\bibfnamefont{G.}~\bibnamefont{Semerjian}},
  \bibinfo{journal}{J. Stat. Mech.} p. \bibinfo{pages}{P04004}
  (\bibinfo{year}{2008}).

\bibitem[{\citenamefont{Ricci-Tersenghi}()}]{Fede}
\bibinfo{author}{\bibfnamefont{F.}~\bibnamefont{Ricci-Tersenghi}},
  \bibinfo{note}{private communication}.

\bibitem[{\citenamefont{Seitz et~al.}(2005)\citenamefont{Seitz, Alava, and
  Orponen}}]{SeitzAlava05}
\bibinfo{author}{\bibfnamefont{S.}~\bibnamefont{Seitz}},
  \bibinfo{author}{\bibfnamefont{M.}~\bibnamefont{Alava}}, \bibnamefont{and}
  \bibinfo{author}{\bibfnamefont{P.}~\bibnamefont{Orponen}},
  \bibinfo{journal}{J. Stat. Mech.} p. \bibinfo{pages}{P06006}
  (\bibinfo{year}{2005}).

\bibitem[{mak()}]{makewff}
\bibinfo{note}{Found at: http://www.cs.rochester.edu/~kautz/walksat/}.

\bibitem[{\citenamefont{Bayardo and Pehousek}(2000)}]{BayardoPehousek00}
\bibinfo{author}{\bibfnamefont{R.~J.} \bibnamefont{Bayardo}} \bibnamefont{and}
  \bibinfo{author}{\bibfnamefont{J.~D.} \bibnamefont{Pehousek}}, in
  \emph{\bibinfo{booktitle}{Proc. 17th AAAI}} (\bibinfo{publisher}{AAAI Press},
  \bibinfo{address}{Menlo Park, California}, \bibinfo{year}{2000}), pp.
  \bibinfo{pages}{157--162}.

\bibitem[{\citenamefont{Monasson
  et~al.}(1999{\natexlab{b}})\citenamefont{Monasson, Zecchina, Kirkpatrick,
  Selman, and Troyansky}}]{MonassonZecchina99b}
\bibinfo{author}{\bibfnamefont{R.}~\bibnamefont{Monasson}},
  \bibinfo{author}{\bibfnamefont{R.}~\bibnamefont{Zecchina}},
  \bibinfo{author}{\bibfnamefont{S.}~\bibnamefont{Kirkpatrick}},
  \bibinfo{author}{\bibfnamefont{B.}~\bibnamefont{Selman}}, \bibnamefont{and}
  \bibinfo{author}{\bibfnamefont{L.}~\bibnamefont{Troyansky}},
  \bibinfo{journal}{Rand. Struct. Algo.} \textbf{\bibinfo{volume}{15}},
  \bibinfo{pages}{414} (\bibinfo{year}{1999}{\natexlab{b}}).

\bibitem[{\citenamefont{Wilson}(2002)}]{Wilson02}
\bibinfo{author}{\bibfnamefont{D.~B.} \bibnamefont{Wilson}},
  \bibinfo{journal}{Rand. Struct. Algo.} \textbf{\bibinfo{volume}{21}},
  \bibinfo{pages}{182} (\bibinfo{year}{2002}).

\bibitem[{\citenamefont{Leone et~al.}(2001)\citenamefont{Leone,
  Ricci-Tersenghi, and Zecchina}}]{LeoneRicci01}
\bibinfo{author}{\bibfnamefont{M.}~\bibnamefont{Leone}},
  \bibinfo{author}{\bibfnamefont{F.}~\bibnamefont{Ricci-Tersenghi}},
  \bibnamefont{and} \bibinfo{author}{\bibfnamefont{R.}~\bibnamefont{Zecchina}},
  \bibinfo{journal}{J. Phys. A} \textbf{\bibinfo{volume}{34}},
  \bibinfo{pages}{4615} (\bibinfo{year}{2001}).

\bibitem[{\citenamefont{Ardelius and Aurell}(2006)}]{ArdeliusAurell06}
\bibinfo{author}{\bibfnamefont{J.}~\bibnamefont{Ardelius}} \bibnamefont{and}
  \bibinfo{author}{\bibfnamefont{E.}~\bibnamefont{Aurell}},
  \bibinfo{journal}{Phys. Rev. E} \textbf{\bibinfo{volume}{74}},
  \bibinfo{pages}{037702} (\bibinfo{year}{2006}).

\bibitem[{\citenamefont{Parisi}(2003)}]{Parisi03}
\bibinfo{author}{\bibfnamefont{G.}~\bibnamefont{Parisi}}
  (\bibinfo{year}{2003}), \bibinfo{note}{arXiv:cs/0301015}.

\bibitem[{\citenamefont{Chavas et~al.}(2005)\citenamefont{Chavas, Furtlehner,
  M\'ezard, and Zecchina}}]{ChavasFurtlehner05}
\bibinfo{author}{\bibfnamefont{J.}~\bibnamefont{Chavas}},
  \bibinfo{author}{\bibfnamefont{C.}~\bibnamefont{Furtlehner}},
  \bibinfo{author}{\bibfnamefont{M.}~\bibnamefont{M\'ezard}}, \bibnamefont{and}
  \bibinfo{author}{\bibfnamefont{R.}~\bibnamefont{Zecchina}},
  \bibinfo{journal}{J. Stat. Mech.} p. \bibinfo{pages}{P11016}
  (\bibinfo{year}{2005}).

\end{thebibliography}

\end{document}